\documentclass[12pt]{article}
\usepackage{cite}
\usepackage{color}
\usepackage{graphicx}
\usepackage{amsmath}
\usepackage{amssymb}
\usepackage{xspace}

\def\baselinestretch{1.2}
\newcommand{\be}{\begin{equation}}
\newcommand{\ee}{\end{equation}}
\newcommand{\beq}{\begin{eqnarray}}
\newcommand{\eeq}{\end{eqnarray}}

\newcommand{\gone}[1]{{}}

\begin{document}
\begin{titlepage}
\begin{flushright}
MAD-TH-07-02
\end{flushright}

\vfil\

\begin{center}

{\Large{\bf Non-commutativity and Open Strings Dynamics\\ in Melvin Universes}}

\vfil

Danny Dhokarh, Akikazu Hashimoto, and 
Sheikh Shajidul Haque

\vfil

Department of Physics, University of Wisconsin, Madison, WI 53706

\vfil

\end{center}

\begin{abstract}
\noindent  We compute the Moyal phase factor for open strings ending on D3-branes wrapping a NSNS Melvin universe in a decoupling limit explicitly using world sheet formalism in cylindrical coordinates.
\end{abstract}
\vspace{0.5in}

\end{titlepage}
\renewcommand{\baselinestretch}{1.05}  

Melvin Universe is an exact axially symmetric solution of Einstein
gravity in a background with magnetic flux \cite{Melvin:1963qx}. 
It arises naturally as a Kaluza-Klein reduction of twisted flat space
\be ds^2 = -dt^2 + d\vec x^2 + dr^2 + r^2 (d \varphi + \eta dz)^2 + dz^2 \ , \label{twist}\ee
along the coordinate $z$. The twist is parameterized by variable
$\eta$. The fact that $z \sim z + 2 \pi R$ is periodic makes the twist
deformation physical.

Melvin universes has a natural embedding in string theory
\cite{Dowker:1993bt,Dowker:1994up,Behrndt:1995si}. Simply embed
(\ref{twist}) in 11-dimensional supergravity. Reducing along $z$ gives
rise to a background in type IIA string theory with a background of
magnetic RR 2-form field strength.

Along similar lines, one can embed (\ref{twist}) in type IIA
supergravity and T-dualize along $z$. This gives rise to a background
in type IIB string theory
\beq ds^2 & = & -dt^2 + d \vec x^2 + dr^2 + {r^2 d \varphi^2 \over 1 + \eta^2 r^2} + {1 \over 1 + \eta^2 r^2} d\tilde z^2 \cr
B & = & {\eta r^2 \over 1 + \eta^2 r^2} d \varphi \wedge d \tilde z \cr
e^{\phi} & = & \sqrt{1 \over 1 + \eta^2 r^2} \cr
\tilde z &=& \tilde z + 2 \pi \tilde R, \qquad \tilde R  = {\alpha' \over R} \ ,  \label{melvinbg}
\eeq
with an axially symmetric magnetic NSNS 3-form field strength in the
background. String theories in backgrounds like (\ref{melvinbg}) are
very special in that the world sheet theory is exactly solvable
\cite{Russo:1994cv,Tseytlin:1994ei,Russo:1995tj,Tseytlin:1995fh,Russo:1995aj,Russo:1995ik}. Quantization
of open strings in Melvin backgrounds have also been studied and was
shown to be exactly solvable \cite{Dudas:2001ux,Takayanagi:2001aj} as
well.

Embedding D-branes in Melvin universes can give rise to interesting
field theories in the decoupling limit. A D3-brane extended along $t$,
$\tilde z$, and two of the $\vec x$ coordinates gives rise to a
non-local field theory known as the ``dipole'' theory
\cite{Bergman:2000cw,Bergman:2001rw}. Orienting the D3-brane to be
extended along the $t$, $r$, $\varphi$, and $\tilde z$ coordinates, on
the other hand, gives rise to a non-commutative gauge theory with a
non-constant non-commutativity parameter\footnote{The first explicit
construction of models of this type is
\cite{Hashimoto:2002nr}.}\cite{Hashimoto:2004pb,Hashimoto:2005hy}. These
are field theories, whose Lagrangian \cite{Hashimoto:2005hy} is
expressed most naturally using the deformation quantization formula of
Kontsevich\footnote{General construction of non-commutative field
theory on curved space-time with non-constant non-commutativity
parameter, arising from D-branes in non-vanishing $H$ field
background, and their relation to the deformation quantization formula
of Kontsevich, was first discussed in \cite{Cornalba:2001sm}.}
\cite{Kontsevich:1997vb}. Field theories arising as a decoupling
limits of various orientations of D-branes in Melvin and related
closed string backgrounds along these lines\footnote{The S-dual NCOS theories with non-constant non-commutativity parameter was studied in  \cite{Cai:2002sv,Cai:2006tda}.}  were tabulated and
classified in Table 1 of \cite{Hashimoto:2004pb}.\footnote{More
recently, a novel non-local field theory, not included in the
classification of \cite{Hashimoto:2004pb}, was discovered
\cite{Ganor:2006ub,Ganor:2007qh}.}

To show that the decoupled field theory is a non-commutative field
theory, the authors of \cite{Hashimoto:2004pb} presented the following
arguments:
\begin{itemize}
\item The application of Seiberg-Witten formula\footnote{The normalization of $B$ field is such that $B_{{\rm Hashimoto-Thomas}}= 2 \pi \alpha' B_{{\rm Seiberg-Witten}}$.} \cite{Seiberg:1999vs}
\be (G + {\theta \over 2 \pi \alpha'})^{\mu \nu} = [(g + B)_{\mu \nu}]^{-1}   \label{swmap} \ee
to the closed string background (\ref{melvinbg}) gives the following open string metric and the non-commutativity parameter
\beq G_{\mu \nu} dx^\mu dx^\nu &=& -dt^2 + dr^2 + r^2 d \varphi^2 + d
z^2 \cr \theta^{\varphi z} & = & 2 \pi \alpha' \eta \
 \label{openmetric} \eeq
which are finite if $\alpha'$ is scaled to zero keeping $\Delta = \alpha' \eta$ fixed.

\item Solution of the classical equations of motion of an open string traveling freely on the D3-brane with angular momentum $J$ has a dipole structure whose size is given by\cite{Hashimoto:2004pb}
\be L = \theta^{\varphi z} J \ . \ee

\end{itemize}

Another suggestive argument is the similarity between $\alpha'
\rightarrow 0$ limit of critical string theory and the boundary
Poisson sigma-model \cite{Cattaneo:1999fm} as was pointed out, e.g.,
in \cite{Baulieu:2001fi}. As was emphasized in \cite{Baulieu:2001fi},
however, the two theories are not to be understood as being equivalent
or continuously connected. This apparent similarity therefore does not
constitute a proof that the decoupled theory is a non-commutative
field theory.

A physical criteria for non-commutativity is the Moyal-like phase
factor in scattering amplitudes.  Scattering amplitudes of open
strings ending on a D-brane can be computed along the lines reviewed
in \cite{Hashimoto:1996bf}.  In the case of the constant
non-commutativity parameter, one can show very explicitly that
\be \langle e^{i p^1 x(\tau_1)} e^{i p^2 x(\tau_2)} \ldots e^{i p^n x(\tau_n)}\rangle_{G,\theta} = e^{-{i \over 2} \sum_{n > m} p_i^n \theta^{ij} p_j^m \epsilon(\tau_n-\tau_m)}
\langle e^{i p^1 x(\tau_1)} e^{i p^2 x(\tau_2)} \ldots e^{i p^n x(\tau_n)}\rangle_{G,\theta=0}\label{master}\ee
which implies that the scattering amplitudes receive corrections in
the form of the Moyal phase factor
\cite{Chu:1998qz,Schomerus:1999ug,Seiberg:1999vs}. The goal of this
article is to derive the analogous statement (\ref{melvinMoyal}) for
the model of \cite{Hashimoto:2004pb,Hashimoto:2005hy}. Once
(\ref{melvinMoyal}) is established in polar coordinates, the
connection to Kontsevich formula follows from performing a change of
coordinates to the rectangular coordinate system and a non-local field
redefinition as is described in
\cite{Cerchiai:2003yu,Hashimoto:2005hy}.

\begin{figure}
\centerline{\includegraphics{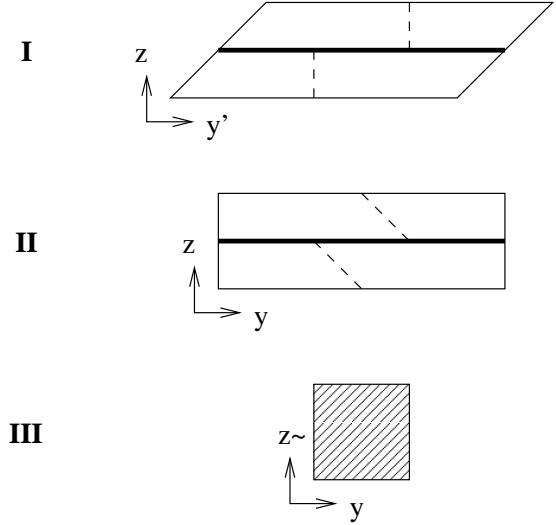}}
\caption{In {\bf I} and {\bf II}, the thick line denotes a D2-brane,
and the dotted line is the minimum energy configuration of the open
strings ending on the D2-branes.  The {\bf I} and {\bf II} are related
by coordinate transformation $y' = y + \eta z$.  {\bf III} is the
T-dual of {\bf II}, and the shaded region in {\bf III} denotes a
D3-brane.\label{figa}}
\end{figure}

A useful first step in this exercise is to reproduce the master
relation (\ref{master}) in a slightly different formalism than what
was used in \cite{Seiberg:1999vs}.  Let us begin by constructing the
closed string background as follows. Start with flat space
\be ds^2 = dy'^2 + d\tilde z^2 \ ,  \ee
where $y$ and $\tilde z$ are compactified with period $L = 2 \pi R$. Then,
\begin{enumerate}
\item[{\bf I}]
T-dualize along the $z$ direction so that the metric becomes 
\be ds^2 = dy'^2 + d z^2 \ .  \ee
\item[{\bf II}] Twist by shifting the coordinates $y' = y + \eta z$
\be ds^2 = (dy+ \eta dz)^2 + d z^2 \ . \ee
\item[{\bf III}] T-dualize on $z$ so that
\be ds^2 = {1 \over 1 + \eta^2}(dy^2 + d \tilde z^2), \qquad B = {\eta \over 1 + \eta^2} dy \wedge d \tilde z \ . \ee
The open string metric associated to this background is
\be G_{\mu \nu} dx^\mu dx^\nu = dy^2 + d \tilde z^2, \qquad \theta^{y \tilde z} = 2 \pi \Delta^2 \ee
if we scale
\be \Delta^2 = \alpha' \eta \ . \ee
\end{enumerate}
The transformation of the coordinates and the orientation of the branes are illustrated in figure \ref{figa}.  This sequence of dualities was referred to as the ``Melvin shift twist'' in \cite{Hashimoto:2004pb}.

The approach of \cite{Seiberg:1999vs} was to work directly in the
duality frame {\bf III}, but one can just as easily work in a
framework which makes the T-duality between duality frame {\bf II}
and {\bf III} manifest, by working with a sigma model of the form 
\be S= {1 \over 2 \pi \alpha'} \int d \sigma_1  d \sigma_2 \left[{1 \over 2} \delta^{ab} \left( \partial_a y \partial_b y + 2 \eta \partial_a y V_b + (1 + \eta^2) V_a V_b\right) + i \epsilon^{ab} \partial_a \tilde z  V_b \right]\label{sig1}\ee
where we have chosen to work in conformal gauge in Eucledian
signature. This action utilizes the Bushar's formulation of T-duality
\cite{Buscher:1987qj}. To see this more explicitly, consider integrating out the field $\tilde z$. This imposes the constraint
\be d V = 0 \rightarrow V_a = \partial_a z \ee
which brings the action (\ref{sig1}) into the form
\be S= {1 \over 2 \pi \alpha'} \int d \sigma_1 d \sigma_2 \left[{1 \over 2} \delta^{ab} \left( \partial_a y \partial_b y + 2 \eta \partial_a y \partial_b z + (1 + \eta^2) \partial_a z \partial_b z \right)  \right]\label{sig2}\ee
which is the sigma model for the duality frame {\bf II}. On the other hand, integrating out $V$ first gives rise to a sigma model of the form
\be S= {1 \over 2 \pi \alpha'} \int d \sigma_1 d \sigma_2 \left[\left( {1 \over 1 + \eta^2}\right) {1 \over 2} \delta^{ab} \left( \partial_a y \partial_b y + \partial_a \tilde z \partial_b \tilde z \right)  + i \left({\eta \over 1 + \eta^2}\right)\epsilon^{ab} \partial_a y \partial_b \tilde z \right]\label{sig3}\ee
which is the string action for the duality frame {\bf III}.

In extracting non-commutative gauge theory as a decoupling limit, we
are interested in embedding a D-brane extended along the $y$ and
$\tilde z$ coordinates in the duality frame {\bf III}.  We must
therefore take the sigma model to be defined on a Riemann surface with
one boundary, which we take to be the upper half plane.  It is also
necessary to impose the appropriate boundary condition for all of the
world sheet fields. We impose the boundary condition which is free
along the $y$ direction and Dirichlet along the $z$ direction:
\be \left.\partial_n y(\sigma,\bar \sigma)+ \eta V_n(\sigma,\bar\sigma)\right|_{\partial \Sigma} = 0 \label{bc0} \ , \ee
\be \left. V_t \right|_{\partial \Sigma}  = \left. \partial_t z \right|_{\partial \Sigma} = 0 \label{bc1} \ . \ee
Using the equation of motion from the variation of $V_a$ field
\be \eta \partial_b y + (1 + \eta^2) V_b + i \epsilon_{ab} \partial_a \tilde z = 0 \, \label{bc2}\ee
and (\ref{bc1}), we infer 
\be  \partial_n \tilde z - i \eta \partial_t  y = 0 \label{bc3}\ .  \ee
The boundary conditions (\ref{bc0}) and (\ref{bc3}) are precisely the
boundary condition imposed in the analysis of \cite{Seiberg:1999vs}.

In order to complete the derivation of (\ref{master}), we add a source term
\be e^{- S_{source}} = \prod_n e^{i k_{yn} y(\sigma_n,\bar \sigma_n) + i k_{zn} \tilde z (\sigma_n,\bar\sigma_n)}  =
e^{\sum_n (i k_{yn} y(\sigma_n,\bar \sigma_n) + i k_{zn} \tilde z (\sigma_n,\bar \sigma_n))} \ee
to the action (\ref{sig1}).  Integrating out the $V$ fields and
bringing the sigma model (\ref{sig1}) into duality frame {\bf III}
would lead to identical computation as what was described in
\cite{Seiberg:1999vs} to derive (\ref{master}).  We will show below
that the same conclusion can be reached using a slightly different
argument which turns out to easily generalize to the case of Melvin
deformed theories \cite{Hashimoto:2004pb,Hashimoto:2005hy}.

The approach we take here is to go to the duality frame {\bf I}. This brings the sigma model (\ref{sig1}) to a simpler form
\be S= {1 \over 2 \pi \alpha'} \int d \sigma_1 d \sigma_2 \left[{1 \over 2} \delta^{ab} \left( \partial_a y' \partial_b y' +  \partial_a z \partial_b z\right)  \right]\label{sig4}\ . \ee
The $\tilde z$ field in the vertex operator now plays the role of a
disorder operator of the dual field $z$. It has the effect of shifting
the Dirichlet boundary condition, incorporating the fact that strings
are stretched along the $z$ direction in frames {\bf I} and {\bf
II}. Also, the fact that the periodicity in $(y',z)$ coordinate system
are twisted
\be (y',z) = (y'+\eta L n, z+L n) \ee
requires an insertion of a disorder operator for the
$y'(\sigma,\bar\sigma)$ field as well. We therefore find that the
source term has the form
\be e^{- S_{source}} = \prod_n e^{i k_{yn} y'(\sigma_n,\bar \sigma_n) + i \eta k_{zn} \tilde y'(\sigma_n, \bar \sigma_n)- i \eta k_{yn} z(\sigma_n,\bar \sigma_n) + i k_{zn} \tilde z (\sigma_n)}  \ . \ee
The boundary condition is now simply Neumann for $y'$ 
\be \left.\partial_n y'(\sigma, \bar \sigma) = 0\right|_{\partial \Sigma} \ , \label{neu} \ee
and Dirichlet for $z$ 
\be \left.\partial_t z(\sigma, \bar \sigma) = 0 \right|_{\partial \Sigma} \ . \label{dir} \ee
In this form, $y'$ and the $z$ sector decouple, allowing their
correlators to be computed separately. In order to compute the
correlation functions involving order and disorder operators with
boundary conditions (\ref{neu}) and (\ref{dir}), it is convenient to
decompose the fields into holomorphic and anti holomorphic parts
\be y'(\sigma, \bar \sigma) = y'(\sigma) + \bar y'(\bar \sigma)\ , \qquad
\tilde y'(\sigma, \bar \sigma) = y'(\sigma) - \bar y'(\bar \sigma)\ , \ee
\be z(\sigma, \bar \sigma) = z(\sigma) + \bar z(\bar \sigma)\ , \qquad
\tilde z'(\sigma, \bar \sigma) = z(\sigma) - \bar z(\bar \sigma) \ . \ee
Their correlation functions are given by
\be \langle y'(\sigma_1) y'(\sigma_2) \rangle  = -{1 \over 2} \alpha' \log(\sigma_1 - \sigma_2) \ee
\be \langle \bar y'(\sigma_1) \bar y'(\sigma_2) \rangle
=  -{1 \over 2} \alpha'\log(\bar \sigma_1 - \bar \sigma_2) \ee
\be \langle \bar y'(\bar \sigma_1) y'(\sigma_2) \rangle =
 - {1 \over 2} \alpha'\log(\bar \sigma_1 - \sigma_2) \ee
\be \langle z(\sigma_1) z(\sigma_2) \rangle  = 
-{1 \over 2}\alpha'\log(\sigma_1 -
 \sigma_2) \ee
\be \langle \bar z (\bar \sigma_1) \bar z(\bar \sigma_2) \rangle
=  -{1 \over 2}\alpha' \log(\bar \sigma_1 - \bar \sigma_2) \ee
\be \langle \bar z(\bar \sigma_1) z(\sigma_2) \rangle =
 {1 \over 2}\alpha'\log(\bar \sigma_1 - \sigma_2), \ee
from which we infer
\be  \langle  y'(\sigma_1,\bar \sigma_1) y'(\sigma_2,\bar \sigma_2) \rangle=  -{1 \over 2} \alpha'(
\log(\sigma_1- \sigma_2) +
\log(\sigma_1- \bar \sigma_2) +
\log(\bar \sigma_1- \sigma_2) +
\log(\bar \sigma_1- \bar \sigma_2) ) \label{coor1}  \ee
\be \langle \tilde y'(\sigma_1,\bar \sigma_1) y'(\sigma_2,\bar \sigma_2) \rangle
=  - {1 \over 2} \alpha'  (
\log(\sigma_1- \sigma_2) +
\log(\sigma_1- \bar \sigma_2) -
\log(\bar \sigma_1- \sigma_2) -
\log(\bar \sigma_1- \bar \sigma_2) ) \label{coor2} \ee
\be  \langle \tilde y'(\sigma_1,\bar\sigma_1) \tilde y'(\sigma_2,\bar \sigma_2) \rangle
 =  - {1 \over 2}  \alpha'  (
\log(\sigma_1- \sigma_2) -
\log(\sigma_1- \bar \sigma_2) -
\log(\bar \sigma_1- \sigma_2) +
\log(\bar \sigma_1- \bar \sigma_2) ) \label{coor3} \ee
\be  \langle \tilde z(\sigma_1,\bar \sigma_1) \tilde z(\sigma_2,\bar \sigma_2) \rangle=  -{1 \over 2} \alpha'(
\log(\sigma_1- \sigma_2) +
\log(\sigma_1- \bar \sigma_2) +
\log(\bar \sigma_1- \sigma_2) +
\log(\bar \sigma_1- \bar \sigma_2) ) \label{coor4} \ee
\be \langle z(\sigma_1,\bar \sigma_1) \tilde z(\sigma_2,\bar \sigma_2) \rangle
=  - {1 \over 2} \alpha'  (
\log(\sigma_1- \sigma_2) +
\log(\sigma_1- \bar \sigma_2) -
\log(\bar \sigma_1- \sigma_2) -
\log(\bar \sigma_1- \bar \sigma_2) ) \label{coor5} \ee
\be  \langle z(\sigma_1,\bar\sigma_1) z(\sigma_2,\bar \sigma_2) \rangle
 =  - {1 \over 2}  \alpha'  (
\log(\sigma_1- \sigma_2) -
\log(\sigma_1- \bar \sigma_2) -
\log(\bar \sigma_1- \sigma_2) +
\log(\bar \sigma_1- \bar \sigma_2) ) \label{coor6} \ . \ee

In terms of these correlation functions, one can easily show that
\beq \lefteqn{\langle {\cal O}(\sigma_1, \bar \sigma_1) {\cal
O}(\sigma_2, \bar \sigma_2) \rangle}\label{corr1} \\
& =& {1 \over 2} \alpha' (k_{y1} k_{y2} + k_{z1} k_{z2}) (\log(\sigma_1 -
\sigma_2) + \log(\sigma_1 - \bar \sigma_2) + \log(\bar \sigma_1 -
\sigma_2) + \log(\bar \sigma_1 - \bar \sigma_2)) \nonumber \cr
&& - \eta \alpha' (k_{y1} k_{z2} - k_{y2} k_{z1}) (\log(\sigma_1 -
\bar \sigma_2) -  \log(\bar \sigma_1 - \sigma_2)) \nonumber\cr
&&
+{1 \over 2}\eta^2  \alpha' (k_{y1} k_{y2} + k_{z1} k_{z2}) (\log(\sigma_1 -
\sigma_2) - \log(\sigma_1 - \bar \sigma_2) - \log(\bar \sigma_1 -
\sigma_2) + \log(\bar \sigma_1 - \bar \sigma_2)) \nonumber
\eeq
for 
\be {\cal O}_n(\sigma_n,\bar \sigma_n) = 
i k_{yn} y'(\sigma_n,\bar \sigma_n) + i \eta k_{zn} \tilde y'(\sigma_n, \bar \sigma_n)- i \eta k_{yn} z(\sigma_n,\bar \sigma_n) + i k_{zn} \tilde z (\sigma_n,\bar \sigma_n) \ . 
\ee
When these operators are pushed toward the boundary
\be \sigma \rightarrow \tau + 0^+ i \ , \ee
the correlation function (\ref{corr1}) reduces to
\be \langle {\cal O}(\tau_1) {\cal
O}(\tau_2) \rangle
 = 2 \alpha' (k_{y1} k_{y2} + k_{z1} k_{z2}) \log(\tau_1 -
\tau_2)
 -  \pi i   \eta \alpha' (k_{y1} k_{z2} - k_{y2} k_{z1}) \epsilon(\tau_2 - \tau_1) \label{OOcorr}\ee
where $\epsilon(\tau)$, following the notation of \cite{Seiberg:1999vs}, is a function which takes the values $\pm 1$ depending on the sign of $\tau$. The term of order $\eta^2$ vanishes in this limit.  From these
results, we conclude that
\be \langle \prod e^{O_n(\tau_n)} \rangle = e^{\sum_{m < n} \langle O_m(\tau_m) O_n(\tau_n) \rangle} \label{OOope}\ee
from which the main statement (\ref{master})  follows immediately.

Finally, let us discuss the generalization of (\ref{master}) to
D3-brane embedded into Melvin universe background (\ref{melvinbg})
along the lines of \cite{Hashimoto:2004pb,Hashimoto:2005hy}. We will
consider the simplest case of embedding (\ref{melvinbg}) into bosonic
string theory. For the Melvin universe background (\ref{melvinbg}), it
is convenient to prepare a vertex operator that corresponds to
tachyons in cylindrical basis
\beq V(\nu,m,\vec k) &=&
\int d k_1 \,  d k_2\,  \delta(\nu^2 - k_1^2 - k_2^2) e^{i m \theta} e^{i  k_1 x_1(\sigma,\bar \sigma) + k_2 x_2(\sigma,\bar \sigma) + \vec k \vec x(\sigma,\bar \sigma)} \cr
& = & e^{i \vec k \vec x(\sigma,\bar \sigma)} J_\nu(r(\sigma,\bar \sigma)) e^{i m \varphi(\sigma,\bar \sigma)}  \label{polarV} \eeq
where 
\be r^2 = x_1^2+x_2^2, \qquad \varphi = \arg(x_1 + i x_2), \qquad \theta = \arg(k_1 + i k_2) \ . \ee
As long as $\vec k^2 + \nu^2$ are taken to satisfy the on-shell
condition of the tachyon, (\ref{polarV}) is linear combination of
operators of boundary conformal dimension 1, and must itself be an
operator of boundary conformal dimension one. Such construction of
vertex operator as a linear superposition is similar in spirit to what
was considered in \cite{Liu:2002ft,Liu:2002kb}.
\be S= {1 \over 2 \pi \alpha'} \int d \sigma_1 d \sigma_2 \left[{1 \over 2} \delta^{ab} \left( \partial_a r \partial_b r + r^2 \partial_a \varphi \partial_b \varphi + 2 \eta r^2 \partial_a \varphi  V_b + (1 + \eta^2 r^2) V_a V_b\right) + i \epsilon^{ab} \partial_a \tilde z  V_b \right]\label{melsig1}\ee
on the upper half plane. Integrating out $\tilde z$ brings this action to the form appropriate for the analogue of {\bf II}
\be S= {1 \over 2 \pi \alpha'} \int d \sigma_1 d \sigma_2 \left[{1 \over 2} \delta^{ab} \left( \partial_a r \partial_b r + r^2 \partial_a \varphi \partial_b \varphi + 2 \eta r^2 \partial_a \varphi  \partial_b z  + (1 + \eta^2 r^2) \partial_a z \partial_b z\right)  \right]\label{melsig1a}\ . \ee

The vertex operators can be represented as a source term
\be e^{-S_{source}} = \prod_n J_{v_n}(r(\sigma_n,\bar \sigma_n)) e^{i m_n \varphi(\sigma_n,\bar \sigma_n) + i k_{zn} \tilde z (\sigma_n, \bar \sigma_n)}  \ee
where $\tilde z$ is a disorder operator. Now, if we let
\be \varphi'(\sigma,\bar \sigma)   = \varphi(\sigma,\bar \sigma) + \eta z(\sigma,\bar \sigma) \ , \ee
the action becomes
\be S= {1 \over 2 \pi \alpha'} \int d \sigma_1 d \sigma_2 \left[{1 \over 2} \delta^{ab} \left( \partial_a r \partial_b r + r^2 \partial_a \varphi' \partial_b \varphi' +  \partial_a z \partial_b z \right) \right]\label{melsig2}\ee
with 
\be e^{-S_{source}} = \prod_n J_{v_n}(r(\sigma_n,\bar \sigma_n)) e^{{\cal O}_n}  \ee
and
\be {\cal O}_n = 
i m_n \varphi'(\sigma_n, \bar \sigma_n)+ i \eta k_{zn} \tilde \varphi'(\sigma_n,\bar \sigma_n) -i \eta m_n z(\sigma_n,\bar \sigma_n) + i k_{zn} \tilde z (\sigma_n, \bar \sigma_n) \label{melvinO} \ee
where
\be \tilde \varphi'(\sigma,\bar \sigma) \ee
is the disorder field for $\varphi'$ satisfying the relation
\be \partial^a \tilde \varphi' = i \epsilon^{ab} r^2 \partial_b \varphi' \ee
which follows naturally from the Busher rule applied to the $\varphi$ fields.

This time, the problem is slightly complicated by the fact that
$(r,\varphi')$ sector is interacting. It is still the case that $(\varphi',z)$
sector, for some fixed $r(\sigma,\bar \sigma)$, is non-interacting. We
will exploit this fact and do the path integral in the order where we
integrate over $\varphi'$ and $z$ first.
The two point function of $\varphi'$ formally has the form
\be \langle \varphi'(\sigma_1, \bar \sigma_1)\varphi'(\sigma_2, \bar \sigma_2) \rangle = (\partial r^2(\sigma,\bar \sigma) \partial)^{-1} \ .  \ee
Then, it follows that
\be \langle \varphi'(\sigma_1, \bar \sigma_1)\partial^a \tilde \varphi'(\sigma_2, \bar \sigma_2) \rangle = i \epsilon^{ab} (\partial^b)^{-1}  
\ee
from which it follows 
\be \langle \tilde \varphi'(\sigma_1,\bar \sigma_1) \varphi'(\sigma_2,\bar \sigma_2) \rangle
= - {1 \over 2} \alpha'  (
\log(\sigma_1- \sigma_2) +
\log(\sigma_1- \bar \sigma_2) -
\log(\bar \sigma_1- \sigma_2) -
\log(\bar \sigma_1- \bar \sigma_2) ) \label{coor7} \ee
in complete analogy with (\ref{coor2}). The correlator (\ref{coor7})
tells us that while the field-field correlator $\langle \varphi' \varphi'
\rangle$ is complicated and $r$ dependent, the field/disorder field
correlator $\langle \tilde \varphi' \varphi' \rangle$ stays simple and
topological.

We can then proceed to compute the analogue of (\ref{OOcorr}) and
(\ref{OOope}) for the operator (\ref{melvinO}) in the $(\varphi',z)$
sector.  While we do not explicitly compute the $\langle \tilde \varphi'
\tilde \varphi' \rangle$ correlator which appear at order $\eta^2$ in
(\ref{OOcorr}), it is clear that the boundary condition forces this
term to vanish as was the case in the earlier example.  The term of
order $\eta$ in the exponential can be made to take the Moyal-like
form
\be e^{{i \over 2}  \sum_{a < b} (2 \pi \Delta) (m_a k_{zb} - k_{za} m_b) \epsilon(\tau_b-\tau_a)}  \label{melvinMoyal}\ee
which is finite in the scaling limit $\alpha' \rightarrow 0$ with
\be \eta = {\Delta \over \alpha'} \ee
keeping $\Delta$ finite. This is precisely the scaling considered in
\cite{Hashimoto:2004pb,Hashimoto:2005hy}. The dependence on $r(\sigma,
\bar \sigma)$ drops out for this term of order $\eta$, allowing us to
further path integrate over this field trivially, with the only effect
of $\eta$ being the overall phase factor (\ref{melvinMoyal}).  This
establishes that the decoupled theory of D-branes in Melvin universes
considered in \cite{Hashimoto:2004pb,Hashimoto:2005hy} has an
effective dynamics which includes the Moyal-like phase factor
involving the angular momentum quantum number $m$ and the momentum
$k_z$. In Cartesian coordinates, this Moyal phase corresponds to a
position dependent non-commutativity
\cite{Hashimoto:2004pb,Hashimoto:2005hy}. This analysis extends
straight forwardly to other simple models of position dependent
non-commutativity, such as\footnote{Using the terminology of
\cite{Hashimoto:2004pb}.} the ``Melvin Null Twist''
\cite{Hashimoto:2002nr} and ``Null Melvin Twist''
\cite{Dolan:2002px}. It would be interesting to extend this analysis
to superstrings and to consider the scattering of states other than
the open string tachyon.

\section*{Acknowledgements}
We would like to thank
I.~Ellwood and 
O.~Ganor
for discussions.
This work was supported in part by the DOE grant DE-FG02-95ER40896 and
funds from the University of Wisconsin.

\bibliography{melvin}\bibliographystyle{utphys}

\providecommand{\href}[2]{#2}\begingroup\raggedright\begin{thebibliography}{10}

\bibitem{Melvin:1963qx}
M.~A. Melvin, ``Pure magnetic and electric geons,'' {\em Phys. Lett.} {\bf 8}
  (1964)
65--70.

\bibitem{Dowker:1993bt}
F.~Dowker, J.~P. Gauntlett, D.~A. Kastor, and J.~H. Traschen, ``Pair creation
  of dilaton black holes,'' {\em Phys. Rev.} {\bf D49} (1994) 2909--2917,
\href{http://www.arXiv.org/abs/hep-th/9309075}{{\tt hep-th/9309075}}.

\bibitem{Dowker:1994up}
F.~Dowker, J.~P. Gauntlett, S.~B. Giddings, and G.~T. Horowitz, ``On pair
  creation of extremal black holes and Kaluza-Klein monopoles,'' {\em Phys.
  Rev.} {\bf D50} (1994) 2662--2679,
\href{http://www.arXiv.org/abs/hep-th/9312172}{{\tt hep-th/9312172}}.

\bibitem{Behrndt:1995si}
K.~Behrndt, E.~Bergshoeff, and B.~Janssen, ``Type II Duality Symmetries in Six
  Dimensions,'' {\em Nucl. Phys.} {\bf B467} (1996) 100--126,
\href{http://www.arXiv.org/abs/hep-th/9512152}{{\tt hep-th/9512152}}.

\bibitem{Russo:1994cv}
J.~G. Russo and A.~A. Tseytlin, ``Constant magnetic field in closed string
  theory: An Exactly solvable model,'' {\em Nucl. Phys.} {\bf B448} (1995)
  293--330,
\href{http://www.arXiv.org/abs/hep-th/9411099}{{\tt hep-th/9411099}}.

\bibitem{Tseytlin:1994ei}
A.~A. Tseytlin, ``Melvin solution in string theory,'' {\em Phys. Lett.} {\bf
  B346} (1995) 55--62,
\href{http://www.arXiv.org/abs/hep-th/9411198}{{\tt hep-th/9411198}}.

\bibitem{Russo:1995tj}
J.~G. Russo and A.~A. Tseytlin, ``Exactly solvable string models of curved
  space-time backgrounds,'' {\em Nucl. Phys.} {\bf B449} (1995) 91--145,
\href{http://www.arXiv.org/abs/hep-th/9502038}{{\tt hep-th/9502038}}.

\bibitem{Tseytlin:1995fh}
A.~A. Tseytlin, ``Exact solutions of closed string theory,'' {\em Class. Quant.
  Grav.} {\bf 12} (1995) 2365--2410,
\href{http://www.arXiv.org/abs/hep-th/9505052}{{\tt hep-th/9505052}}.

\bibitem{Russo:1995aj}
J.~G. Russo and A.~A. Tseytlin, ``Heterotic strings in uniform magnetic
  field,'' {\em Nucl. Phys.} {\bf B454} (1995) 164--184,
\href{http://www.arXiv.org/abs/hep-th/9506071}{{\tt hep-th/9506071}}.

\bibitem{Russo:1995ik}
J.~G. Russo and A.~A. Tseytlin, ``Magnetic flux tube models in superstring
  theory,'' {\em Nucl. Phys.} {\bf B461} (1996) 131--154,
\href{http://www.arXiv.org/abs/hep-th/9508068}{{\tt hep-th/9508068}}.

\bibitem{Dudas:2001ux}
E.~Dudas and J.~Mourad, ``D-branes in string theory Melvin backgrounds,'' {\em
  Nucl. Phys.} {\bf B622} (2002) 46--72,
\href{http://www.arXiv.org/abs/hep-th/0110186}{{\tt hep-th/0110186}}.

\bibitem{Takayanagi:2001aj}
T.~Takayanagi and T.~Uesugi, ``D-branes in Melvin background,'' {\em JHEP} {\bf
  11} (2001) 036,
\href{http://www.arXiv.org/abs/hep-th/0110200}{{\tt hep-th/0110200}}.

\bibitem{Bergman:2000cw}
A.~Bergman and O.~J. Ganor, ``Dipoles, twists and noncommutative gauge
  theory,'' {\em JHEP} {\bf 10} (2000) 018,
\href{http://www.arXiv.org/abs/hep-th/0008030}{{\tt hep-th/0008030}}.

\bibitem{Bergman:2001rw}
A.~Bergman, K.~Dasgupta, O.~J. Ganor, J.~L. Karczmarek, and G.~Rajesh,
  ``Nonlocal field theories and their gravity duals,'' {\em Phys. Rev.} {\bf
  D65} (2002) 066005,
\href{http://www.arXiv.org/abs/hep-th/0103090}{{\tt hep-th/0103090}}.

\bibitem{Hashimoto:2002nr}
A.~Hashimoto and S.~Sethi, ``Holography and string dynamics in time-dependent
  backgrounds,'' {\em Phys. Rev. Lett.} {\bf 89} (2002) 261601,
\href{http://www.arXiv.org/abs/hep-th/0208126}{{\tt hep-th/0208126}}.

\bibitem{Hashimoto:2004pb}
A.~Hashimoto and K.~Thomas, ``Dualities, twists, and gauge theories with
  non-constant non-commutativity,'' {\em JHEP} {\bf 01} (2005) 033,
\href{http://www.arXiv.org/abs/hep-th/0410123}{{\tt hep-th/0410123}}.

\bibitem{Hashimoto:2005hy}
A.~Hashimoto and K.~Thomas, ``Non-commutative gauge theory on D-branes in
  Melvin universes,'' {\em JHEP} {\bf 01} (2006) 083,
\href{http://www.arXiv.org/abs/hep-th/0511197}{{\tt hep-th/0511197}}.

\bibitem{Cornalba:2001sm}
L.~Cornalba and R.~Schiappa, ``Nonassociative star product deformations for
  D-brane worldvolumes in curved backgrounds,'' {\em Commun. Math. Phys.} {\bf
  225} (2002) 33--66,
\href{http://www.arXiv.org/abs/hep-th/0101219}{{\tt hep-th/0101219}}.

\bibitem{Kontsevich:1997vb}
M.~Kontsevich, ``Deformation quantization of Poisson manifolds, I,'' {\em Lett.
  Math. Phys.} {\bf 66} (2003) 157--216,
\href{http://www.arXiv.org/abs/q-alg/9709040}{{\tt q-alg/9709040}}.

\bibitem{Cai:2002sv}
R.-G. Cai, J.-X. Lu, and N.~Ohta, ``NCOS and D-branes in time-dependent
  backgrounds,'' {\em Phys. Lett.} {\bf B551} (2003) 178--186,
\href{http://www.arXiv.org/abs/hep-th/0210206}{{\tt hep-th/0210206}}.

\bibitem{Cai:2006tda}
R.-G. Cai and N.~Ohta, ``Holography and D3-branes in Melvin universes,'' {\em
  Phys. Rev.} {\bf D73} (2006) 106009,
\href{http://www.arXiv.org/abs/hep-th/0601044}{{\tt hep-th/0601044}}.

\bibitem{Ganor:2006ub}
O.~J. Ganor, ``A new Lorentz violating nonlocal field theory from string
  theory,''
\href{http://www.arXiv.org/abs/hep-th/0609107}{{\tt hep-th/0609107}}.

\bibitem{Ganor:2007qh}
O.~J. Ganor, A.~Hashimoto, S.~Jue, B.~S. Kim, and A.~Ndirango, ``Aspects of
  Puff Field Theory,''
\href{http://www.arXiv.org/abs/hep-th/0702030}{{\tt hep-th/0702030}}.

\bibitem{Seiberg:1999vs}
N.~Seiberg and E.~Witten, ``String theory and noncommutative geometry,'' {\em
  JHEP} {\bf 09} (1999) 032,
\href{http://www.arXiv.org/abs/hep-th/9908142}{{\tt hep-th/9908142}}.

\bibitem{Cattaneo:1999fm}
A.~S. Cattaneo and G.~Felder, ``A path integral approach to the Kontsevich
  quantization formula,'' {\em Commun. Math. Phys.} {\bf 212} (2000) 591--611,
\href{http://www.arXiv.org/abs/math.qa/9902090}{{\tt math.qa/9902090}}.

\bibitem{Baulieu:2001fi}
L.~Baulieu, A.~S. Losev, and N.~A. Nekrasov, ``Target space symmetries in
  topological theories. I,'' {\em JHEP} {\bf 02} (2002) 021,
\href{http://www.arXiv.org/abs/hep-th/0106042}{{\tt hep-th/0106042}}.

\bibitem{Hashimoto:1996bf}
A.~Hashimoto and I.~R. Klebanov, ``Scattering of strings from D-branes,'' {\em
  Nucl. Phys. Proc. Suppl.} {\bf 55B} (1997) 118--133,
\href{http://www.arXiv.org/abs/hep-th/9611214}{{\tt hep-th/9611214}}.

\bibitem{Chu:1998qz}
C.-S. Chu and P.-M. Ho, ``Noncommutative open string and D-brane,'' {\em Nucl.
  Phys.} {\bf B550} (1999) 151--168,
\href{http://www.arXiv.org/abs/hep-th/9812219}{{\tt hep-th/9812219}}.

\bibitem{Schomerus:1999ug}
V.~Schomerus, ``D-branes and deformation quantization,'' {\em JHEP} {\bf 06}
  (1999) 030,
\href{http://www.arXiv.org/abs/hep-th/9903205}{{\tt hep-th/9903205}}.

\bibitem{Cerchiai:2003yu}
B.~L. Cerchiai, ``The Seiberg-Witten map for a time-dependent background,''
  {\em JHEP} {\bf 06} (2003) 056,
\href{http://www.arXiv.org/abs/hep-th/0304030}{{\tt hep-th/0304030}}.

\bibitem{Buscher:1987qj}
T.~H. Buscher, ``Path integral derivation of quantum duality in nonlinear sigma
  models,'' {\em Phys. Lett.} {\bf B201} (1988)
466.

\bibitem{Liu:2002ft}
H.~Liu, G.~W. Moore, and N.~Seiberg, ``Strings in a time-dependent orbifold,''
  {\em JHEP} {\bf 06} (2002) 045,
\href{http://www.arXiv.org/abs/hep-th/0204168}{{\tt hep-th/0204168}}.

\bibitem{Liu:2002kb}
H.~Liu, G.~W. Moore, and N.~Seiberg, ``Strings in time-dependent orbifolds,''
  {\em JHEP} {\bf 10} (2002) 031,
\href{http://www.arXiv.org/abs/hep-th/0206182}{{\tt hep-th/0206182}}.

\bibitem{Dolan:2002px}
L.~Dolan and C.~R. Nappi, ``Noncommutativity in a time-dependent background,''
  {\em Phys. Lett.} {\bf B551} (2003) 369--377,
\href{http://www.arXiv.org/abs/hep-th/0210030}{{\tt hep-th/0210030}}.

\end{thebibliography}\endgroup

\end{document}